\documentclass[twocolumn,showpacs,preprintnumbers,amsmath,amssymb,prl]{revtex4-1}

\usepackage{mathrsfs}
\usepackage{graphicx}
\usepackage{dcolumn}
\usepackage{bm}
\usepackage{amsmath}
\usepackage{amsfonts}
\begin{document}

\title{Discovery of a Weyl Semimetal in non-Centrosymmetric Compound TaAs}

\author{L. X. Yang$^{\dagger,1,2,3}$, Z. K. Liu$^{\dagger,4,5}$, Y. Sun$^{\dagger,6}$, H. Peng$^2$, H. F. Yang$^{2,7}$, T. Zhang$^{1,2}$, B. Zhou$^{2,3}$, Y. Zhang$^3$, Y. F. Guo$^2$, M. Rahn$^2$, P. Dharmalingam$^2$,  Z. Hussain$^3$, S. –K. Mo$^3$, C. Felser$^{6,8}$, B. Yan$^{5,6}$ and Y. L. Chen$^{\ast,2,1,4,5}$}
\affiliation{
$^1$ State Key Laboratory of Low Dimensional Quantum Physics, Department of Physics, Tsinghua University, Beijing 100084, P. R. China\\
$^2$ Physics Department, Oxford University, Oxford, OX1 3PU, UK\\
$^3$ Advanced Light Source, Lawrence Berkeley National Laboratory, Berkeley, CA 94720, USA\\
$^4$ Diamond Light Source, Harwell Science and Innovation Campus, Fermi Ave, Didcot, Oxfordshire, OX11 0QX, UK\\
$^5$ School of Physical Science and Technology, ShanghaiTech University, Shanghai 200031, China\\
$^6$ Max Planck Institute for Chemical Physics of Solids, D-01187 Dresden, Germany\\
$^7$ State Key Laboratory of Functional Materials for Informatics, SIMIT, Chinese Academy of Sciences, Shanghai 200050, China\\
$^8$ Institut f{\"u}r Anorganische Chemie und Analytische Chemie, Johannes Gutenberg-Universtit{\"a}t, 55099 Mainz, Germany
}

\begin{abstract}
  Three-dimensional (3D) topological Weyl semimetals (TWSs) represent a novel state of quantum matter with unusual electronic structures that resemble both a "3D graphene" and a topological insulator by possessing pairs of Weyl points (through which the electronic bands disperse linearly along all three momentum directions) connected by topological surface states, forming the unique "Fermi-arc" type Fermi-surface (FS). Each Weyl point is chiral and contains half of the degrees of freedom of a Dirac point, and can be viewed as a magnetic monopole in the momentum space. Here, by performing angle-resolved photoemission spectroscopy on non-centrosymmetric compound TaAs, we observed its complete band structures including the unique "Fermi-arc" FS and linear bulk band dispersion across the Weyl points, in excellent agreement with the theoretical calculations \cite{wenghm2015,huang2015}. This discovery not only confirms TaAs as the first 3D TWS, but also provides an ideal platform for realizing exotic physical phenomena (e.g. negative magnetoresistance, chiral magnetic effects and quantum anomalous Hall effect) which may also lead to novel future applications.
\end{abstract}

\date{March 17th, 2015}

\maketitle

\section{Introduction}
The discovery of quantum materials with non-trivial topological electronic structures such as topological insulators, topological crystalline insulators and Dirac semimetals \cite{qixl2011,hasan2010,chen2009,fu2011,dziawa2012,wangzj2012,liuzk2014sci}, has recently ignited worldwide interest due to their rich scientific implications and broad application potentials \cite{qixl2011,hasan2010,chen2009,fu2011,dziawa2012,wangzj2012,liuzk2014sci}. Although being the subject of condensed matter physics, the research on topological quantum matter has benefited from the connection to other fields of physics, such as high energy physics, by the introduction of Dirac and Majorana fermions into the electronic spectra of crystals. Recently, another intriguing particle - the Weyl fermion - which was also originally introduced in high energy physics (e.g. as a description of neutrinos), is proposed to have its counterpart in solid state physics \cite{wan2011}, leading to a new type of topological quantum matter, the topological Weyl semimetals (TWSs) \cite{wenghm2015,huang2015,wan2011,xug2011,burkov2011}.

\begin{figure*}[!]
\includegraphics[width=\textwidth]{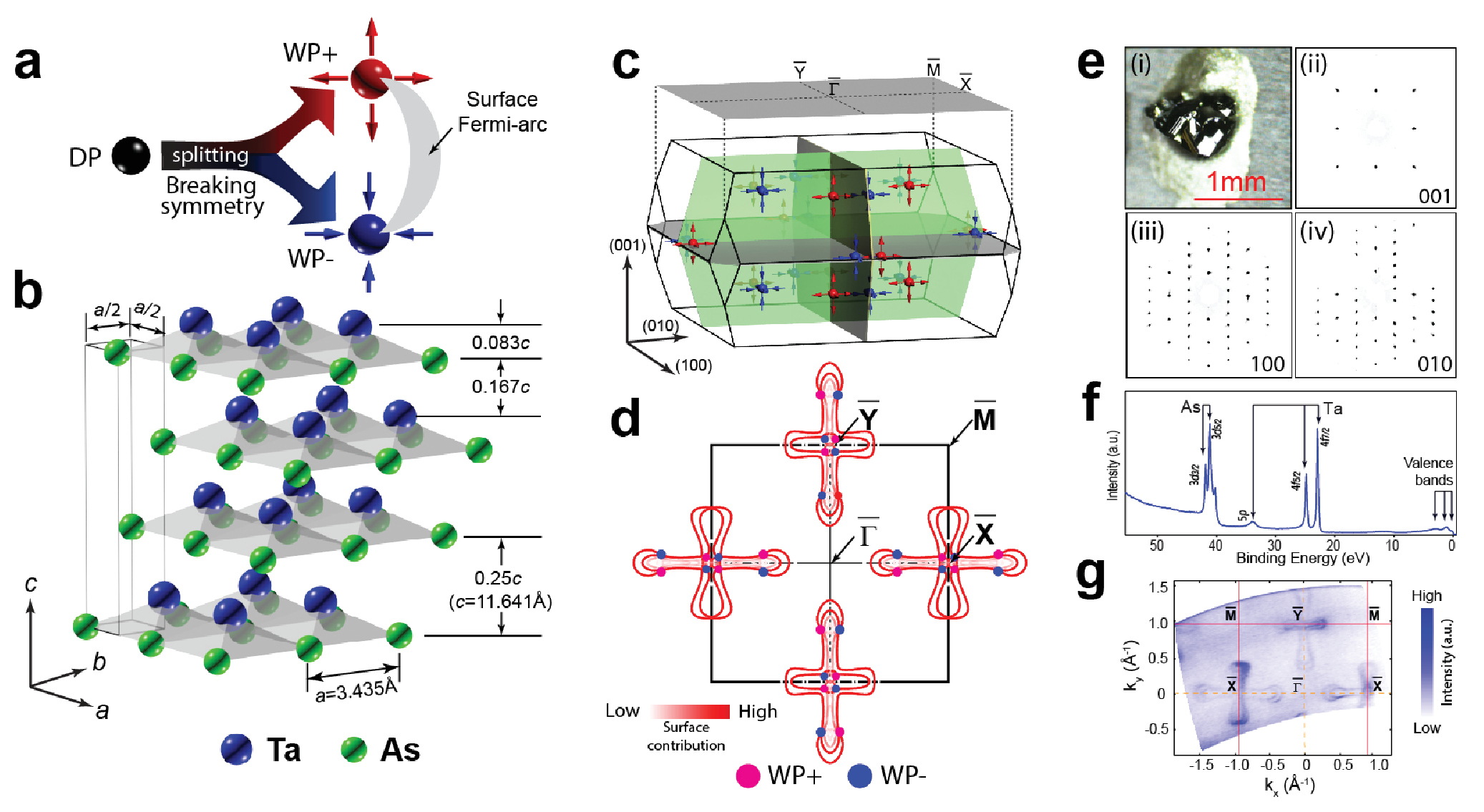}
  \caption{a. Illustration of the splitting of a Dirac point (DP). A DP can be splitted into a pair of Weyl points with opposite chirality (marked as WP+ and WP- and behave as the "source" or "sink" of Berry curvature) by breaking time-reversal or inversion symmetry (see text for discussions). The two Weyl points are connected by the Fermi-arc type of FS formed by the topological surface states. b. Crystal structure of TaAs, showing the ...A-B-C-D... stacking of TaAs layers. c. Schematic of the bulk and (001) surface Brillouin zones (BZs) of TaAs. Twelve pairs of Weyl points are predicted in each BZ, with four pairs at the kz=0 and $\pm$1.16 $\pi$/c planes, respectively. d. FS from ab initio calculations are plotted on the (001) surface BZ with the (projected) Weyl points overlapped, showing characteristic Fermi-arc FS geometry. Color bar shows the surface contribution of the FS (white/0$\%$ to red/100$\%$), same in Fig. 2-4. e. (i) Image of the TaAs single crystal with flat cleavage plane used for ARPES measurements. (ii-iv) X-ray diffraction patterns of the TaAs crystal from different crystalline directions. f. Core level photoemission spectrum clearly shows the characteristic As 3\emph{d}, Ta 5\emph{p} and 4\emph{f} peaks. g. Broad FS map confirms the (001) cleavage plane and the lattice constant in b, the uneven intensity of the FS at different BZs results from the matrix element effect.
}

\end{figure*}

\begin{figure*}
\includegraphics[width=\textwidth]{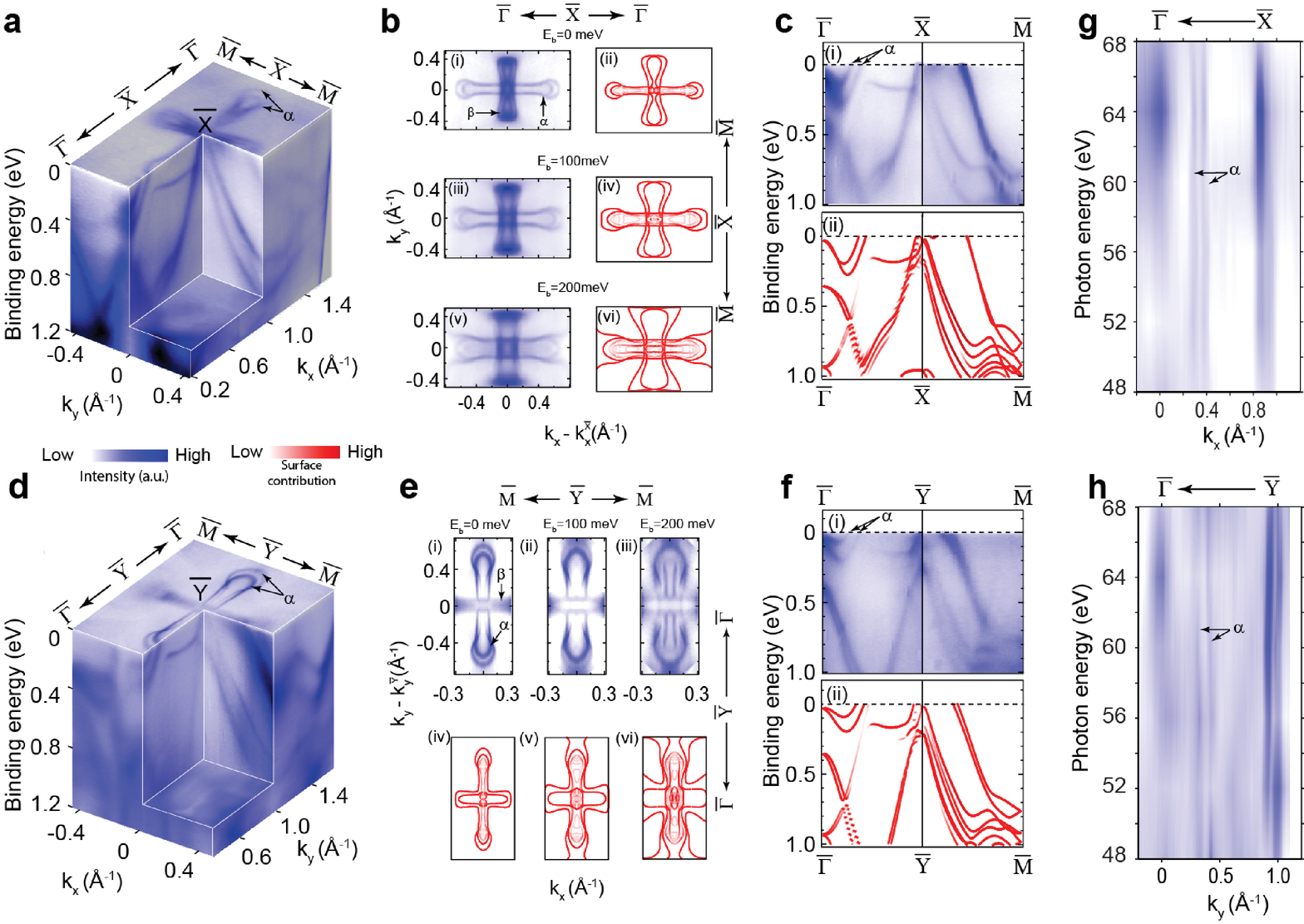}
  \caption{a. 3D intensity plot of the photoemission spectra around the $\bar{X}$ point, showing the band dispersions and the resulted FSs. The spoon-like $\alpha$-FSs are marked. b. Comparison of three constant energy contours at different binding energies between experiments and ab initio calculations shows excellent agreements. The spoon/bowtie-like $\beta$-FSs (see text) are marked. For better comparison with calculations, the experiment plot has been symmetrized with respect to k$_y$=0 plane according to the crystal symmetry (same in e below). c. Comparison of experiment and calculated dispersions along high symmetry directions: $\bar{\Gamma}$ - $\bar{X}$ - $\bar{M}$. d-f, same as a-c, around the $\bar{Y}$ point of the surface BZ. g, h. Photon-energy-dependent ARPES measurement plot of photoemission intensities at E$_F$ along the high symmetry $\bar{\Gamma}$ - $\bar{X}$ (g) and $\bar{\Gamma}$ - $\bar{Y}$  (h) directions as a function of photon energy (48-68eV), showing the k$_z$ dispersion of different bands. The $\alpha$ bands in c, f are labelled.
}

\end{figure*}

\begin{figure*}
\includegraphics[width=\textwidth]{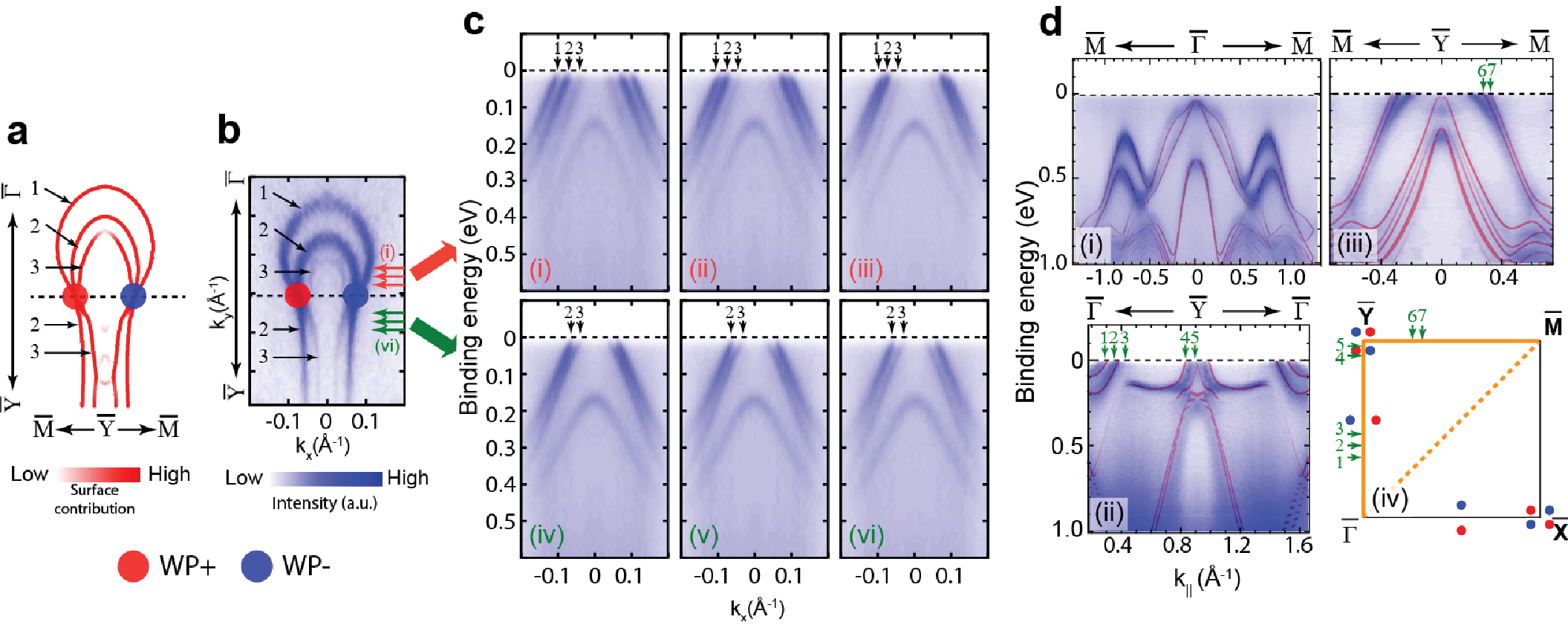}
  \caption{a. Calculated FS geometry with fine k-space grid around the $\bar{Y}$ point of the BZ showing the Fermi-arc FS (see text and ref \cite{wenghm2015,huang2015}). Different FS pieces are labelled as 1-3, and the dished lines connects the two Weyl points for reference. b. FSs measured by ARPES with high resolution show excellent agreement with those in a. Three arrows above and below the dashed line indicate the measurements positions in c. c, Three band dispersions measured above and below the dashed line in b (with the locations indicated by the red and green arrows in b, respectively). Band dispersions corresponding to the 1-3 FS pieces in b are also labelled as 1-3, respectively. d. (i-iii) Dispersions from ARPES measurements along different high symmetry directions overlapped by the ab initio calculations, showing excellent agreement. The Fermi-crossing locations 1-7 are marked by green arrows in panels (ii) and (iii), with the corresponding positions in the surface BZ also marked in panel (iv). (iv), The surface BZ with the Weyl points and the Fermi-crossings marked.
}

\end{figure*}

\begin{figure*}
\includegraphics[width=\textwidth]{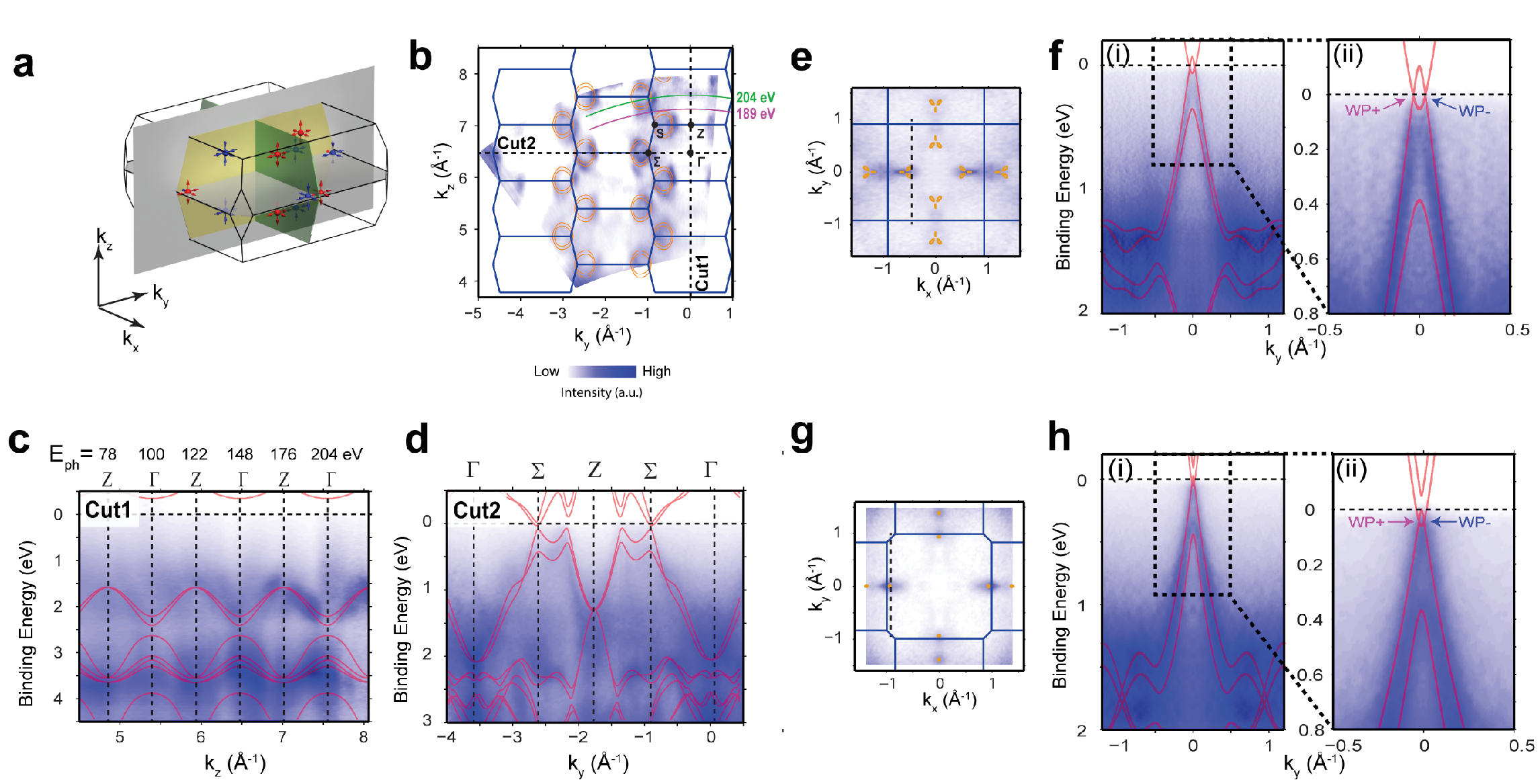}
  \caption{a. Schematic illustration of the measurement k$_y$-k$_z$ plane (vertical grey plane) of the intensity plot in b, Weyl points are also shown. b. Photoemission intensity plot of the k$_y$-k$_z$ plane (k$_x$=0, the grey plane in a). The integration window is from E$_F$-100meV to E$_F$. Overlapped yellow curves are calculated bulk bands' constant energy contour at the center energy of the intensity map (E$_F$ - 50meV), showing excellent agreement. Green and magenta curves indicate the k$_z$ momentum locations probed by 204 eV and 189 eV photons, respectively. The two dashed lines marked as "cut1" and "cut2" indicate the momentum direction of the two band dispersions shown in c, d. c, d. Bulk band dispersions along two high-symmetry directions, indicated as cut1 and cut2 in b, respectively. The calculated band dispersions (red curves) are overlaid. e.. FS map of the k$_x$-k$_y$ plane using 189 eV photons with integration window from E$_F$-20meV to E$_F$+20meV, with the calculated FS overlapped (orange pockets). Black dashed line indicates the measurement direction of the band dispersion in f, and the blue lines indicate the k$_x$-k$_y$ BZ at k$_z$=1.16$\pi$/c (in the reduced BZ). For better comparison with calculations, the experiment plot has been symmetrized with respect to k$_x$=0 and k$_y$=0 planes according to the crystal symmetry (same in f-h). The uneven spectra intensity between k$_x$=0 and k$_y$=0 directions are due to the matrix element effect. f. Broad band dispersion (i) with the zoomed-in plot (ii) across the Weyl points (along the direction indicated by the dashed line in e(ii)). g, h. Same as e, f for photon energy of 204 eV (through the Weyl points at k$_z$=0 (in the reduced BZ)).
}

\end{figure*}

A TWS exhibits unique band structures that resemble to both a "3D graphene" and a topological insulator. On one hand, the bulk conduction and valence bands of a TWS touch linearly at pairs of discrete points - the Weyl points, through which the bands disperse linearly along all three momentum directions (thus is a 3D analogue of graphene); as Weyl points of opposite chirality can be either the "source" or "sink" of Berry curvature \cite{wenghm2015,huang2015,wan2011} (Fig. 1a), they can also be viewed as magnetic monopoles in the momentum space. On the other hand, in a TWS, there exist unique surface "Fermi-arcs"\cite{wenghm2015,huang2015,wan2011,xug2011,burkov2011} (Fig. 1a), or unclosed Fermi-surfaces originated from the topological surface states (similar to those in topological insulators) that start and end at the Weyl points of opposite chirality (Fig. 1a). The unique bulk Weyl fermions and the surface Fermi-arcs can give rise to many unusual physical phenomena, such as negative magnetoresistance, chiral magnetic effects, quantum anomalous Hall effect, novel quantum oscillations (in magneto-transport) and quantum interference (in tunneling spectroscopy) \cite{zyuzin2012,liucx2013,landsteiner2014,potter2014,hosur2012}.

In principle, a TWS can be realized by breaking either the time reversal symmetry (TRS) or inversion symmetry \cite{burkov2011,halasz2012} of a topological Dirac semimetals (TDS) recently discovered \cite{liuzk2014sci,liuzk2014nmat,neupane2014,borisenko2014}. This way, the bulk Dirac point in a TDS can be split into two Weyl points (Fig. 1a), thus realizing the TWS state. However, this method does not generate intrinsic TWSs and the need for large external magnetic field or mechanical strain requires complicated instrumentation and experiment setup (thus limits the use and application of these materials). Under this circumstance, the pursuit of intrinsic materials with spontaneously broken symmetry has become the focus of current research. Up to date, several candidates with naturally broken TRS have been proposed (e.g. Y$_2$Ir$_2$O$_7$ \cite{wan2011} and HgCr$_2$Se$_4$ \cite{xug2011}), however, none of them have been experimentally confirmed, leaving the existence of the TWS elusive.

Recently, another type of TWS candidates with naturally broken inversion symmetry were proposed in several compound families, including finely tuned solid solutions LaBi$_{1-x}$Sb$_x$Te$_3$, LuBi$_{1-x}$Sb$_x$Te$_3$ \cite{liujp2014} and transition metal monoarsenides/phorsphides (including TaAs, TaP, NbAs and NbP) \cite{wenghm2015}. In this work, by using angle resolved photoemission spectroscopy (ARPES), we systematically studied the electronic structure of single crystal TaAs and observed the unique surface "Fermi-arcs" on its Fermi-surface as well as the linear bulk band dispersions through the Weyl points. The excellent agreement between our experimental band structures and ab initio calculations (including previous theoretical predictions \cite{wenghm2015,huang2015}, clearly establishes that TaAs is a TWS.

\section{Experimental methods}

\textbf{Sample Growth.} Precursor polycrystalline TaAs samples were prepared by mixing high purity ($>$99.99$\%$)  Ta and As elements and the mixture was sealed into a quartz tube under high vacuum, which was sealed into another evacuated tube for extra protection.  First, the vessel was heated to 600$^{\circ}{\rm C}$ at the rate of 50$^{\circ}{\rm C}$/hour, after 10 hours of socking, it was slowly heated to 1050$^{\circ}{\rm C}$ at 30$^{\circ}{\rm C}$/hour rate and kept at this temperature for 24 hours. Finally the vessel was cooled down to room temperature.

With the polycrystalline precursor, the single crystals were grown using chemical vapor transport method in a two zone furnace.  The polycrystalline TaAs powder and 0.46 mg/cm$^3$ of Iodine was loaded into a 24 mm diameter quartz tube and sealed under a vacuum.  The charge part of the tube was kept at 1150$^{\circ}{\rm C}$ and the other end at 1000$^{\circ}{\rm C}$ for three weeks.  The resulting crystals can be as large as 0.5-1 mm in size.

\textbf{Angle resolved photoemission spectroscopy (ARPES) measurements.} ARPES measurements were performed at beamline 10.0.1 of the Advanced Light Source (ALS) at Lawrence Berkeley National Laboratory, USA and BL I05 of the Diamond Light Source (DLS), UK. The measurement pressure was kept below 3x10$^{-11}$/9x10$^{-11}$ Torr in the two facilities, and data were recorded by Scienta R4000 analyzers at 10 K sample temperature. The total convolved energy and angle resolutions were 16meV and 0.2$^{\circ}$. The fresh surface of TaAs for ARPES measurement was obtained by cleaving the TaAs sample in situ along its natural (001) cleavage plane.

\textbf{LDA calculations.} Electronic structures were calculated by the density-functional theory (DFT) method which is implemented in the Vienna ab initio Simulation Package (VASP) \cite{kresse1996}. The core electrons were represented by the projected augmented wave method \cite{bloch1994}. The exchange-correlation was considered in the generalized gradient approximation (GGA) \cite{perdew1996} and spin-orbital coupling (SOC) was included self-consistently. The energy cut off was set to be 300 eV for the plane-wave basis. Experimental lattice parameters were used in the construction of a slab model with seven-unit-cell thick to simulate a surface, in which the top and bottom surface are terminated by As and Ta, respectively. Positions of outermost four atomic layers were fully optimized to consider the surface atomic relaxation. The surface band structures and Fermi surfaces were projected to the first unit cell of the As-terminated side, which fits the experimental band structure well. We adopted 12x12 and 400x400 k-point grids in the charge self-consistent and Fermi surface calculations, respectively.

\section{Results and discussion}

The crystal structure of TaAs is shown in Fig. 1b. There are four TaAs layers in a unit cell along the c-direction,, forming repeating ...-A-B-C-D-… stacking structure \cite{furuseth1965} without inversion symmetry (Fig. 1b). As the distance along c-direction between intra- and inter-plane Ta and As layer is 0.083c (0.966{\AA}) and 0.167c (1.944{\AA}) respectively, the crystal naturally cleaves between adjacent TaAs layers along the (001) plane (Fig. 1b, Fig. 1e(i)), which is ideal for the ARPES measurements. In recent theoretical investigations1,2, TaAs was proposed as a TWS candidate with twelve pairs of Weyl points in each Brillouin zone (BZ) (Fig. 1c), with each pair of Weyl points connected by topologically non-trivial surface states, forming the unique surface ''Fermi-arcs". The characteristic FS of TaAs with the surface ''Fermi-arcs" on the (001) surface from our ab initio calculations is shown in Fig. 1d, in nice agreement with the previous theoretical works \cite{wenghm2015,huang2015}.

High quality TaAs crystals were synthesized for our ARPES measurements (Fig. 1e(i)), showing flat and shiny cleaved surface. The X-ray diffraction along different crystalline orientations (Fig. 1e(ii-iv)) confirmed the crystal structure. The core level photoemission spectrum (Fig. 1f) show sharp characteristic Ta 5\emph{p}, 4\emph{f} and As 3\emph{d} core levels and the broad Fermi-surface mapping (Fig. 1g) illustrate the overall FS topology agreeing with the ab initio calculations in Fig. 1d (fine measurements with more details will be discussed below).

Due to the intrinsic surface sensitivity, ARPES is an ideal tool to study the unusual surface states and search for the unique surface Fermi-arcs FS in TaAs. In Fig. 2, we illustrate the overall FS geometry and the band structure evolution with different binding energy around both $\bar{X}$ and $\bar{Y}$ points of the surface BZ.

The 3D band structure around both $\bar{X}$ and $\bar{Y}$ regions are presented in Fig. 2a and Fig. 2d respectively, which illustrates the FS geometry with related band dispersions. In panel b and e, three constant energy contours at different binding energies are selected to show the band evolution around $\bar{X}$ and $\bar{Y}$ points - both vary from the cross-shape FSs to more complicated shapes at high binding energy. Each set of the cross-shape FSs (Fig. 2b(i) and Fig. 2e(i)) is comprised of two orthogonal sub-sets of FSs, one forms the spoon-like pockets (marked as $\alpha$-FSs) and the other forms the bowtie-shape pockets (marked as $\beta$-FSs) – all of these FSs agree well with our ab initio calculations (presented side-by-side in Fig. 2b and Fig. 2e).

Besides the FS topology, we also studied the band dispersions across the BZ, which are illustrated in Fig.2c and Fig.2f, respectively. The comparison between the measurements and calculations again shows excellent agreements. The surface nature of the bands that form the spoon-like $\alpha$-FSs (Fig. 2a, b(i), d, e(i)) can be verified by the photon energy dependent ARPES measurements \cite{chen2012} (Fig. 2g, h), where the ARPES spectra clearly show no k$_z$-dispersion, in contrast to the bulk band dispersion which we will discuss later in Fig. 4.

After establishing the overall correspondence between the experimental and theoretical band structures, we zoom into the spoon-like $\alpha$-FSs by performing fine ARPES mapping with high resolution to study the detailed FS geometry and search for the unusual surface Fermi-arcs – the unique signature of a TWS.

In Fig. 3a, our ab initio calculations show clear surface Fermi-arc (green curve, marked as FS-1) terminating at the Weyl points (see Fig. 3a inset for clarity), while the other two FS pieces (FS-2, FS-3) extend across the Weyl points. This unusual FS topology was indeed experimentally observed in Fig. 3b, which matches Fig. 3a excellently. The change of the FS pieces across the Weyl points in experiment can also be verified by the band dispersions (Fig. 3c). Evidently, measurements above the Weyl points (Fig. 3c (i-iii)) show three bands dispersing across E$_F$ while there are only two bands crossing E$_F$ below the Weyl points (Fig. 3c (iv-vi)), caused by the termination of FS-1 at the Weyl points.

Interestingly, as a Fermi-arc is an unclosed FS, we can also verify the existence of Fermi-arcs by counting the total number of Fermi-crossings along a closed loop in the BZ that enclosed an odd number of Weyl points - and get an odd number of total Fermi-crossings. We thus choose the $\bar{\Gamma}$ - $\bar{Y}$ -$\bar{M}$ - $\bar{\Gamma}$ loop in a BZ (see Fig. 3d(iv)) which enclosed three Weyl points, including two degenerate ones (red color, at different k$_z$, but projected to the surface BZ at the same location, see  Fig. 1c, d for details) and a singular one (blue color). In Fig. 3d, the counting of the Fermi-crossings along the $\bar{\Gamma}$ - $\bar{Y}$ (panel (ii)), $\bar{Y}$ - $\bar{M}$ (panel (iii)) and $\bar{M}$ - $\bar{\Gamma}$ (panel (i)) yields five, two and zero crossings, respectively. Thus the total Fermi-crossings along the $\bar{\Gamma}$ - $\bar{Y}$ -$\bar{M}$ - $\bar{\Gamma}$ loop is seven (an odd number), which confirms the existence of the Fermi-arcs on the FS of TaAs.

In addition to the unique surface Fermi-arc, we also carried out ARPES measurements (Fig. 4) with high photon energy to investigate the bulk band structure of TaAs. In Fig. 4b, bulk bands with strong k$_z$ dispersion can be clearly seen in the k$_y$-k$_z$ spectra intensity map, agreeing well with our calculation (overlapped on Fig. 4b, note that the Weyl points are not observed here as they are off the k$_x$=0 plane, see Fig. 1c, d). Also, the measured dispersions along high symmetry directions show nice agreement with calculations (Fig. 4c, d).

These excellent agreements between our experiments and calculations allow us to identify the Weyl points predicted lying at k$_z$=$\pm$1.16$\pi$/c and k$_z$=0 planes (Fig. 1c), which can be accessed by using 189 eV (k$_z$=-1.16$\pi$/c in the reduced BZ) and 204 eV (k$_z$=0 in the reduced BZ) photons, respectively (see Fig. 4b). At each photon energy, we first carried out k$_x$-k$_y$ FS mapping (Fig. 4e, g) to locate the in-plane momentum positions of the Weyl points, then measured the band dispersion across them (Fig. 4f, h). Indeed, the measured bulk band dispersion in both cases show clear linear dispersions that match well with our calculations (Fig. 4f, h) again, confirming the existence of Weyl points in the bulk band structure of TaAs.

\section{Conclusion}
The observation of the unique surface Fermi-arc and the bulk Weyl points with linear dispersions, together with the overall agreement of the measurements with the theoretical calculations, establish TaAs as the first TWS experimentally observed. This discovery further opens a door for the exploration of other exotic phenomena associated with the TWS and potential applications due to the ultra-high mobility and unusually large (and none-saturating) magnetoresistance in 3D semimetals recently discovered \cite{liang2015}.

We note that while we were finalizing this manuscript, two other groups also independently studied the compound TaAs and showed the surface Fermi-arcs \cite{suyangxu2015,bqlv2015}.

\begin{acknowledgments}
Y.L.C acknowledge the support from the EPSRC (UK) grant EP/K04074X/1 and a DARPA (US) MESO project (no. N66001-11-1-4105). Advanced Light Source is operated by Department of Energy, Office of Basic Energy Science (contract DE-AC02-05CH11231).

$^{\dagger}$ L.X.Y., Z.K.L., and Y.S. contributed equally to this work.
\end{acknowledgments}

\bibliography{TaAs_arxiv}
\newpage

\end{document}